\documentclass{revtex4}
\usepackage{graphicx}
\usepackage{amsmath,amsfonts,amssymb,amsthm}
\begin{document}
\title{The scalar complex potential of the electromagnetic field}

\author{ Y. Friedman and  S. Gwertzman \\
Jerusalem College of Technology, Israel\\
e-mail: friedman@jct.ac.il}

\begin{abstract}
In this paper, we define a scalar complex potential
$\mathcal{S}$ for an arbitrary electromagnetic field. This
potential is a modification of the two scalar potential functions
introduced by E. T. Whittaker. By use of a complexified
Minkowski space $M$, we decompose the usual Lorentz
group representation on $M$  into a product of two commuting new representations.  These
representations are based on the complex Faraday tensor. For a moving charge  and for any observer, we obtain
a complex dimensionless scalar which is invariant under one of our new representations. The scalar complex potential
is the logarithm of this dimensionless scalar times the charge value. We define a  conjugation  on $M$ which is invariant under our representation. We show that the Faraday tensor is the derivative of the conjugate of the gradient of the complex potential. The real part of the Faraday tensor  coincides with the usual electromagnetic tensor of the field.

The potential $\mathcal{S}$, as a complex-valued function on  space-time, is described as an integral over the distribution of the charges generating the electromagnetic field. This potential is like
a wave function description of the field. If we chose the Bondi tetrad (called also Newman-Penrose basis) as a basis on $M$, the  components of the Faraday vector at each point may be derived from $\mathcal{S}$ by $ F_j=E_j+iB_j={\partial}^\nu (\alpha_j)_\nu^\lambda{\partial}_\lambda \mathcal{S}$, where $(\alpha_j)$ are the known $\alpha$-matrices of Dirac.
This fact indicates that our potential may build a "bridge" between classical and quantum physics.



\end{abstract}
 \maketitle

\section{Introduction}
In general, the electromagnetic  field
tensor ${F}$, expressed by a $4\times 4$  antisymmetric matrix, is used to describe
the electromagnetic field intensity.  This tensor provides a
convenient expression for the Lorentz force and therefore is used to
describe the motion of charged particles in this field. Another way to describe an
electromagnetic  field is by use of the 4-potential. In a chosen gauge, the 4-potential
transforms  as a 4-vector. The electromagnetic  field
tensor is defined as the derivative of the 4-potential.

In 1904 E. T. Whittaker introduced \cite{Whittaker} two scalar potential functions.
He showed that the electromagnetic field can be expressed in terms of the
second derivatives of these functions.  H. S. Ruse improved \cite{Ruse} the result
of Whittaker and showed that these
two functions transform as invariants. We show that it is possible
to combine these two scalar potential functions into one complex-valued function
$\mathcal{S}$ on complexified Minkowski space, which we call the \textit{ scalar complex
potential of the electromagnetic field}.

 Complexified space-time was  also used by Barut \cite{Barut} for introducing a
 classical theory of fields and particles. In \cite{FriedmanPoinc} this space was used for
 different  representations of the Poincar\'{e} group on relativistic phase
space and in \cite{FS} and  \cite{F04} for the description of the motion of a charge in
an electromagnetic field.
It turns out to be easer to express and calculate the scalar complex potential of the electromagnetic
 field by the use of a Bondi tetrad for complexified Minkowski space.
 Such a basis is used widely, see for example \cite{PenroseRindler}, \cite{ODonnell}. The connection of the relativistic phase space to the Bondi tetrad is described in \cite{FriedmanNP}.
We use the complex Faraday tensor $\mathcal{F}$, which was introduced in \cite{Silberstein1} and used in
\cite{Silberstein2} and \cite{FD}. The real part of
this tensor  coincides with the usual electromagnetic tensor of the field. The additional symmetries
of this tensor make it attractive to solve evolution equations and to obtain new representations of the
Lorentz group.

An electromagnetic field is generated by a collection of moving
charges.  Thus, a description of an electromagnetic field can
be obtained by integrating the fields of moving charges. We show that for a moving charge and observer
point in space-time, there is a  complex dimension-less scalar which is invariant under some
representation of the Lorentz group.
The scalar complex potential $\mathcal{S}$ is defined to be the logarithm of this scalar.
We show that the Faraday tensor $\mathcal{F}$ is an appropriately defined second order derivative of the
scalar  complex potential.

In classical mechanics, the negative of the gradient  of a scalar potential equals the force.
 This is true for forces which generate linear acceleration. Such forces are defined by a one-form
(since their line integral gives the work). The derivative of a scalar function
is also a one-form. Hence, the derivative of a potential could be equal to the negative of the
force. But in classical mechanics we also have rotating forces, which are described by
two-forms. Such forces cannot be expressed as derivatives of a scalar potential. In
special relativity, the electromagnetic field is expressed by a two-form. So, it is natural to assume
that a kind of second derivative of a scalar potential  will define
the force. Note that the usual differential form derivative of a gradient is zero.
Therefore, we  define a Lorentz invariant conjugation on the complex space-time. We show that the derivative of the
conjugate of the gradient of the complex potential equals the  Faraday tensor $\mathcal{F}$ of the field.

In Section 2 we recall the definition of the Bondi tetrad on the complexified Minkowski space $M$
and the transformation from the usual basis to it. In Section 3 we introduce a complex
Faraday tensor and define the complex analog of the $curl$ operator on $M$. In Section 4 we
show that this tensor is decomposable if we use Bondi tetrad for $M$.
The Faraday tensor is used in Section 5 to define two Lorentz group
representations on the  complexified space-time. For each representation we define  a conjugation on $M$ which is invariant under it. In section 6 we introduce an invariant scale-free scalar
associated with a null-vector. The logarithm of this scalar becomes the  scalar complex potential
of a moving charge. This potential is introduced in section 7. In Section 8 we calculate the 4-potential of a
moving charge as the conjugate of the gradient of the complex potential. In Section 9 we derive the  Faraday tensor $\mathcal{F}$ of the field of a moving charge as the derivative of the 4-potential. We give an explicit formula for such a field. In Section 10 we show how to extend the scalar complex potential to  arbitrary
electromagnetic field. We show that it satisfies the wave equation and obtain an explicit form of the complex $curl$ operator under this representation.

 \section{Complexified Minkowski space with a Bondi tetrad}

  Let $K$ be an inertial reference frame with basis $\{\mathbf{e}_\mu\}$,
 for $\mu=0,1,2,3,$ and coordinates $(ct,x,y,z)=x^\mu,$ where $c$ denotes the speed of light.
 For the rest of the paper we will use units in which $c=1$ and thus we
will omit $c$ from equations. Greek indices range over $\{0,1,2,3\}$ and Latin indices over $\{1,2,3\}$. The inner product of two 4-vectors is defined, as usual, as
\begin{equation}\label{inner prod mink}
  \mathbf{a}\cdot \mathbf{b}=\eta _{\mu\nu}a^\mu b^\nu ,\;\;\eta _{\mu\nu}=diag(1,-1,-1,-1).
\end{equation}
The space of 4-vectors with this inner product is  Minkowski space-time.
 We now complexify Minkowski space-time by allowing the coefficients $x^\mu$ to be complex numbers. As a space this is equivalent to $\mathbf{C}^4$. We extend the inner product (\ref{inner prod mink}) to a symmetric complex bilinear form. We will call such a space with such a bilinear form on it the \textit{ complexified  Minkowski space} and denote it by $M$.

 One possible interpretation of $M$ is the following: Consider the state space of a zero spin particle which is described by a complex-valued function $\psi$ on the space-time. The gradient operator, describing a generalized momentum, maps the space-time into $M$ since $\nabla\psi\in\mathbf{C}^4$ and the inner product (\ref{inner prod mink}) on $M$ is induced from the inner product on space-time.
 Another possible interpretation of $M$ is:
  An electromagnetic field is described by a two-tensor $F_{\mu\nu}$ defining the action of the field on the 4-velocity $u^\mu$ of a test charge $q$, expressed
 by equation $\frac{m}{q}\frac{dU^\mu}{d\tau}=F^\mu_\nu U^\nu$. The 4-velocity can be considered as a tangent vector to the path of the charge in Minkowski space-time. A typical charge is a collection of electrons. So, we can think of a test charge as a single electron. But a state of an electron describing its evolution depends also on the spin $S$ of the electron, as can be seen from the Stern-Gerlach experiment. The evolution of the spin in the field is given by the BMT equation which, if we assume that the Land$\grave{e}$ factor of the electron to be equal 2, is $\frac{m}{q}\frac{dS^\mu}{d\tau}=F^\mu_\nu S^\nu$. Note that the spin evolution is the same as the evolution of the 4-velocity.
 By complexifying the tangent space of Minkowski space-time, we can define a complex 4-vector $x^\mu=U^\mu+iS^\mu$ that will describe the state of the electron and the space $M$ containing all such state vectors. With this notation the evolution equation will become $ \frac{m}{q}\frac{dx^\mu}{d\tau}=F^\mu_\nu x^\nu$. Note that since $S$ is a pseudo-vector and $U$ is a vector, in order that $x$ will be well defined $i$ must be a  pseudo-scalar which changes its sign with the change of orientation in space.  For other possible interpretations of $M$, see also \cite{F04} and \cite{FriedmanPoinc}.

The \textit{ Bondi null tetrad} ( \textit{BT }, in short), called also Newman-Penrose basis,  on  $M$ is defined by
\[\mathbf{ n}_0=\mathbf{n}=\frac{1}{\sqrt{2}}(\mathbf{e}_0+\mathbf{e}_3),\;
    \mathbf{ n}_1=\bar{\mathbf{m}}=\frac{1}{\sqrt{2}}(\mathbf{e}_1-i\mathbf{e}_2),\;\]
\begin{equation}\label{NPbasisNotaion}
      \mathbf{ n}_2=\mathbf{m}=\frac{1}{\sqrt{2}}(\mathbf{e}_1+i\mathbf{e}_2),\;
      \mathbf{ n}_3=\mathbf{l}=\frac{1}{\sqrt{2}}(\mathbf{e}_0-\mathbf{e}_3).
\end{equation}
For the significance of the
BT see \cite{PenroseRindler}, \cite{ODonnell} and
\cite{FriedmanNP}. Note that application of complex conjugation on $M$, which is equivalent
to replacing $i$ with $-i$, maps the BT into itself, but exchanges $\mathbf{ n}_1$ with
$\mathbf{ n}_2$. This mean that also here $i$ is a pseudo-scalar which with the change
of orientation, that may be expressed by change of the order of the basis vectors, changes its sign.

We will denote  by $y^\mu$ the coordinates of a vector in $M$ with respect
to the BT, meaning $x^\mu \mathbf{e}_\mu=y^\mu \mathbf{n}_\mu.$
Then, the relation between the coordinates is
\begin{equation}\label{r in term q}
   x^0=\frac{1}{\sqrt{2}}(y^0+y^3), \; x^1=\frac{1}{\sqrt{2}}(y^1+y^2),\;
    x^2=i\frac{1}{\sqrt{2}}(y^2-y^1),\;
    x^3=\frac{1}{\sqrt{2}}(y^0-y^3),
\end{equation}
or inversely,
\begin{equation}\label{q in term r}
   y^0=\frac{1}{\sqrt{2}}(x^0+x^3), \;
   y^1=\frac{1}{\sqrt{2}}(x^1+ix^2),\; y^2=\frac{1}{\sqrt{2}}(x^1-ix^2),\;
   y^3=\frac{1}{\sqrt{2}}(x^0-x^3).
\end{equation}

The coordinate transformation could be expressed by the transfer
matrix $L=L_j^k$ given by
\begin{equation}\label{trasfer matrix}
    L=\frac{1}{\sqrt{2}}\left(
        \begin{array}{cccc}
          1 & 0& 0 & 1 \\
           0 & 1 & i & 0\\
          0 & 1 & -i & 0 \\
           1 & 0 & 0 & -1\\
        \end{array}
      \right),\;\; L^{-1}=\frac{1}{\sqrt{2}}\left(
        \begin{array}{cccc}
          1 & 0& 0 & 1 \\
          0 & 1 & 1 & 0 \\
          0 & -i & i & 0 \\
          1 & 0 & 0 & -1 \\
        \end{array}
      \right).
\end{equation}
Thus,
\begin{equation}\label{coord tranf}
    y^\nu=L_\mu^\nu x^\mu,\; \;
    x^\nu=(L^{-1})_\mu^\nu y^\mu\,.
\end{equation}

The metric tensor $\eta$ in the BT is given by
\begin{equation}\label{NPmetric tensor}
    \tilde{\eta}=\left(
        \begin{array}{cccc}
          0 & 0 & 0 & 1 \\
          0 & 0 & -1 & 0 \\
          0 &  -1 & 0 & 0 \\
          1 & 0 & 0& 0 \\
        \end{array}
      \right)= \tilde{\eta}^{-1}.
\end{equation}
The bilinear symmetric scalar product of two  4-vectors  $\mathbf{a}=a^\mu \mathbf{n}_\mu$ and
$\mathbf{b}=b^\mu \mathbf{n}_\mu$ is given by
\[\mathbf{a}\cdot \mathbf{b}=\tilde{\eta} _{\mu\nu}a^\mu b^\nu. \]
This, for example, implies that
\begin{equation}\label{Zero Int NP}
    (\mathbf{a})^2=\mathbf{a}\cdot \mathbf{a}=0\;\;\Leftrightarrow\;\; a^0a^3=a^1a^2 \;\;\Leftrightarrow
    \;\; \frac{a^2}{a^0}=\frac{a^3}{a^1}\;\;\Leftrightarrow\;\; \frac{a^1}{a^0}=\frac{a^3}{a^2}\,.
\end{equation}
In case $a^0=0$, the last identities need to be reversed $\frac{a^0}{a^2}=\frac{a^1}{a^3},\;\frac{a^0}{a^1}=\frac{a^2}{a^3}.$
In these coordinates, the lowering of indices is denoted by $a_\mu=\tilde{\eta}_{\mu\nu} a^\nu$. For example,
\begin{equation}\label{LowerinIndexinNP}
   y_0=y^3,\;\;y_1=-y^2,\;\;y_2=-y^1,\;\;y_3=y^0.
\end{equation}

\section{Operator Representation of the Faraday Vector}
 An electromagnetic field can be defined by an electric field intensity $\mathbf{E}(\mathbf{r},t)$
 and a magnetic field intensity $\mathbf{B}(\mathbf{r},t)$. Equivalently, one can define a complex
3D-vector, called the Faraday vector, as
\begin{equation}\label{FaradayVector}
 \mathbf{F}=\mathbf{E}+i\mathbf{B}
\end{equation}
in order to represent the electromagnetic field. Note that since $i$ is a pseudo-scalar and $\mathbf{B}$
is a pseudo-vector, the expression $i\mathbf{B}$ is a vector which is independent of
the chosen  orientation of the space. The Faraday vector is used
to describe the Lorentz invariant field constants. See, for example, \cite{Landau}.

An alternative way to describe an electromagnetic field is by use of  the 4-potential $A=A_\mu$.
  It is known that the 4-potential  $A=A_\mu$ and the  $j=1,2,3$ components of the electric and the magnetic field intensities $\mathbf{E}, \mathbf{B}$ are connected by
\begin{equation}\label{4-potent andEB}
     E_j=\frac{\partial A_0}{\partial x^j}-\frac{\partial A_j}{\partial x^0}=A_{0,j}-A_{j,0},\;\;
    B_j=\varepsilon_j^{kl}\frac{\partial A_k}{\partial x^l}=\varepsilon_j^{kl}A_{k,l}\,,
\end{equation}
where $\varepsilon^{klj}$ denotes the antisymmetric 3D Levi-Civita tensor, with
$\varepsilon^{123}=-1$. From this we get that the components of the  Faraday vector satisfy
\begin{equation}\label{4-potent andF}
    F_j=E_j+iB_j=\frac{\partial A_0}{\partial x^j}-\frac{\partial A_j}{\partial x^0}+
    i\varepsilon_j^{kl}\frac{\partial A_k}{\partial x^l}.
\end{equation}

We introduce matrices
\begin{equation}\label{JMatrices}
(K_1)_\nu^\mu=\frac{1}{2}\left(
\begin{array}{cccc}
0 & 1 & 0 & 0 \\
1 & 0 & 0 & 0 \\
0 & 0 & 0 & -i \\
0 & 0 & i & 0 \\
\end{array}\right),\;
(K_2)_\nu^\mu=\frac{1}{2}\left(
\begin{array}{cccc}
0 & 0 & 1 & 0 \\
0 & 0 & 0 & i \\
1 & 0 & 0 & 0 \\
0 & -i & 0 & 0 \\
\end{array}\right),\;
(K_3)_\nu^\mu=\frac{1}{2}\left(
\begin{array}{cccc}
0 & 0 & 0 & 1 \\
0 & 0 & -i & 0 \\
0 & i & 0 & 0 \\
1 & 0 & 0 & 0 \\
\end{array}\right)
\end{equation}
and differential operators ${\partial}_\mu=(\frac{\partial}{\partial x^0}
\frac{\partial}{\partial x^1}\frac{\partial}{\partial x^2}\frac{\partial}{\partial x^3})=\frac{\partial}{\partial x^\mu}$ and ${\partial}^\mu=\eta^{\mu\nu}{\partial}_\nu$.
With this notation we can rewrite equation (\ref{4-potent andF}) as
\begin{equation}\label{Fj fromA}
   F_j=-2{\partial}^\nu(K_ j)^\mu_\nu A_\mu  .
\end{equation}

In \cite{FD} we introduce a \textit{complex Faraday tensor} for the description of an electromagnetic
field. This tensor is a complex matrix (mixed tensor)
\begin{equation}\label{FaradayMatrix}
\mathcal{F}^\beta_\alpha =\frac{1}{2} \left(\begin{array}{cccc}
0 & F_1 & F_2 & F_3 \\
F_1 & 0 & -iF_3 & iF_2 \\
F_2 & iF_3 & 0 & -iF_1 \\
F_3 & -iF_2 & iF_1 & 0 \\
\end{array}\right) =\sum_{j=1}^3 F_j  (K_j)^\beta_\alpha\, ,
\end{equation}
with $F_j$ are defined by (\ref{FaradayVector}).

Substituting (\ref{Fj fromA}) into (\ref{FaradayMatrix}) we introduce a differential operator
\begin{equation}\label{Diff of 1form on M}
    (\nabla_c\times)^{\beta\mu}_{\alpha}=\sum_{j=1}^3  (K_j)_{\alpha}^\beta {\partial}^\nu (K_{j})^{\mu}_\nu.
\end{equation}
This operator plays the role of a complex $curl$ on $M$ since
\[ \mathcal{F}_{\alpha}^\beta = - (\nabla_c\times)^{\beta\mu}_{\alpha} A_\mu \,. \]

\section{Complex Faraday tensor in BT}

 We will need the representation of the Faraday tensor in BT. We will denote by $\widetilde{\mathcal{F}}$
the matrix of $\mathcal{F}$ in this representation. By the usual formula of basis transformation we get
\begin{equation}\label{FNP0}
    \widetilde{\mathcal{F}}^\beta_\alpha=L\mathcal{F}^\beta_\alpha L^{-1}=\frac{1}{2}\left(
                                         \begin{array}{cccc}
                                           F_3 & F_1-iF_2 & 0 & 0 \\
                                           F_1+iF_2 & -F_3 & 0 & 0 \\
                                           0 & 0 & F_3 &F_1-iF_2 \\
                                           0 & 0 & F_1+iF_2 & -F_3 \\
                                         \end{array}
                                       \right).
\end{equation}
This show that the Faraday tensor become decomposable in BT. Thus it can be simplified if we introduce the following tensor decomposition.

The tensor decomposition of a $4\times 4$ matrix as a
tensor product of $2\times 2$ matrices is defined by use of the binary representation of numbers. Each
of our indices $\mu=0,1,2,3$ can be considered as a pair of indices $(\mu_0,\mu_1)$ with value in $\mu_k\in\{0,1\}$ by
\[0\longmapsto 00,\;\;1\longmapsto 01,\;\;2\longmapsto 10,\;\;3\longmapsto 11\,.\]
The \textit{tensor decomposition} of a two tensor $D=D_{jk}$ is defined by
\begin{equation}\label{tesnor deconposition}
  D=a\otimes b, \;\;  D_{\mu\nu}=D_{(\mu_0,\mu_1)(\nu_0,\nu_1)}=a_{\mu_0\nu_0}\,b_{\mu_1\nu_1}\,.
\end{equation}
For example, the tensor $\tilde{\eta}$ can be decomposed as $\tilde{\eta}=\left(
                                                     \begin{array}{cc}
                                                       0   & 1 \\
                                                       -1 & 0 \\
                                                     \end{array}
                                                   \right)\otimes \left(
                                                     \begin{array}{cc}
                                                       0   & 1 \\
                                                       -1 & 0 \\
                                                     \end{array}
                                                   \right):=\tilde{\eta}_2\otimes \tilde{\eta}_2\,.$

The following properties of the tensor decomposition can be verified directly from the definition:
\begin{equation}\label{tensor prop lin}
    a\otimes (b+c)=(a\otimes b)+( a\otimes c), \;\;F(a\otimes b)=(Fa)\otimes b=a\otimes Fb
\end{equation}
and
\begin{equation}\label{tensor prop prod}
   ( a\otimes b)(c\otimes d)=ac \otimes bd,
\end{equation}
where $a,b,c,d$ are $2\times 2$ matrices and $F$ is a constant.

With this notation and the properties of the tensor decomposition, we can rewrite (\ref{FNP0}) as
  \begin{equation}\label{FNP}
    \widetilde{\mathcal{F}}^\beta_\alpha=I_2\otimes \sum_jF_j\frac{1}{2}\sigma_j=\sum_j F_j\widetilde{K}_j,
\end{equation}
where $\sigma_j$ denote the usual  Pauli matrices, and $I_2$ is the  $2\times 2$ identity matrix. The matrices
$K_j$ in BT therefore become
\begin{equation}\label{KjinNP}
    (\widetilde{K}_j)^\mu_\nu=I_2\otimes \frac{1}{2}\sigma_j.
\end{equation}
Formula (\ref{Fj fromA}) becomes
\begin{equation}\label{Fj fromA NP}
   F_j=-2\tilde{\partial}^\nu (\tilde{K}_j)^{\mu}_\nu \tilde{A}_\mu =-\tilde{\partial}^\nu (I_2\otimes \sigma_j)^{\mu}_\nu \tilde{A}_\mu,
\end{equation}
where $\tilde{\partial}_\mu=(\frac{\partial}{\partial y^0}
\frac{\partial}{\partial y^1}\frac{\partial}{\partial y^2}\frac{\partial}{\partial y^3})=\frac{\partial}{\partial y^\nu}$, $\tilde{\partial}^\mu=\tilde{\eta}^{\mu\nu}\tilde{\partial}_\nu =\left(
                                                                            \begin{array}{c}
                                                                              \tilde{\partial}_3 \\
                                                                              -\tilde{\partial}_2 \\
                                                                              -\tilde{\partial}_1 \\
                                                                              \tilde{\partial}_0 \\
                                                                            \end{array}
                                                                          \right)
$ and $\tilde{A}_\mu$ is the representation of ${A}_\mu$ in BT.

The analog of the  differential operator $curl$ becomes
\begin{equation}\label{Diff of 1form on Min NP}
    (\widetilde{\nabla}_c\times)^{\beta\mu}_{\alpha}=\sum_{j=1}^3 2( \tilde{K}_j)_{\alpha }^\beta{\tilde{\partial}}^\nu (\tilde{K}_j)^{\mu}_\nu\,
\end{equation}
and
\begin{equation}\label{tensorfroAmuNP}
   \widetilde{\mathcal{F}}_{\alpha\beta}=(\widetilde{\nabla}_c\times)^{\beta\mu}_{\alpha}\tilde{A}_\mu \,.
\end{equation}

The complex Faraday tensor $\mathcal{F}^\beta_\alpha$ defined by (\ref{FaradayMatrix}) is an extension of usual electromagnetic tensor $F^\beta_\alpha$ in the following way:
\begin{equation}\label{Faraday real and compl}
    F^\beta_\alpha=2 Re \mathcal{F}^\beta_\alpha=\mathcal{F}^\beta_\alpha+(\mathcal{F}^*)^\beta_\alpha\, ,
\end{equation}
where $(\mathcal{F}^*)^\beta_\alpha$ denote the complex conjugate of $\mathcal{F}^\beta_\alpha$. In BT,
the tensor $(\mathcal{F}^*)^\beta_\alpha$ becomes
\begin{equation}\label{Fraday adj in NP}
  (\widetilde{\mathcal{F}}^*)^\beta_\alpha=L\mathcal{F}^* L^{-1}=\sum_j\bar{F}_j\frac{1}{2}\sigma^*_j\otimes I_2\,.
\end{equation}
>From this, (\ref{FNP}) and (\ref{tensor prop prod}) it follows that
\begin{equation}\label{commuting f fbar}
 [\mathcal{F},\mathcal{F}^*]=0 \,.
\end{equation}

 \section{Representations of the Lorentz group on $M$}

In the previous section we used both the complex Faraday tensor $\mathcal{F}^\beta_\alpha$ and the usual electromagnetic tensor $F^\beta_\alpha$ to describe the electromagnetic field action on the complexified Minkowski space $M$.  As it was shown in \cite{FriedmanPoinc}, such operators may be considered as elements of the action of Lie algebra of the Lorentz group on $M$.  An electric field, which is connected to acceleration, can be considered as the generator of a boost, and a magnetic field can be considered as the generator of a rotation. Thus, by use of the above operators we can introduce 3 representations of the Lorentz group: $\tilde{\pi}$ based on  $\mathcal{F}^\beta_\alpha$,  $\tilde{\pi}^*$ based on  $(\mathcal{F}^*)^\beta_\alpha$ and $\pi$ based on $F^\beta_\alpha$ on $M$.

We define first the representation $\tilde{\pi}$.
 A generator $\xi $ of a boost $B_k$ in direction $\mathbf{k}=(k_1,k_2,
k_3)$ with $|\mathbf{k}|=1$, can be identified with an electric field $\mathbf{E}=(k_1,k_2,k_3)$. Using the BT
for $M$, from (\ref{FaradayVector}) and (\ref{FNP}), we can represent this generator on $M$ as $\xi=I_2\otimes \sum k_j\frac{1}{2}\sigma_j $. Thus, the representation  $\tilde{\pi}$ of the boost $B_k$ , using (\ref{tensor prop lin}) and (\ref{tensor prop prod}), is
 \[\tilde{\pi}(B_k) =\exp(\xi\psi)=\exp ((I_2\otimes \sum k_j\frac{1}{2}\sigma_j)\psi)= I_2\otimes \exp(\sum k_j\sigma_j\frac{\psi}{2})\] for some constant $\psi$. Using the fact that $(\sum k_j\sigma_j) ^2=I_2$, we get
\begin{equation}\label{rep spin boost}
   \tilde{\pi}(B_k)= I_2\otimes\left(
  \begin{array}{cc}
      \cosh \frac{\psi}{2} + k_3\sinh \frac{\psi}{2} & (k_1-ik_2)\sinh \frac{\psi}{2}  \\
        (k_1+i k_2)\sinh \frac{\psi}{2} & \cosh \frac{\psi}{2} -k_3\sinh \frac{\psi}{2} \\
                      \end{array}
                                    \right)
\end{equation}

A generator $\eta $ of a rotation $R_k$ about the  direction $\mathbf{k}=(k_1,k_2,k_3)$, with $|\mathbf{k}|=1$, can be identified with a magnetic field $\mathbf{B}=(k_1,k_2,k_3)$.Using the BT
for $M$, from (\ref{FaradayVector}) and (\ref{FNP}), we can represent this generator on $M$ as $\eta=I_2\otimes i\sum k_j\frac{1}{2}\sigma_j .$ Thus, the representation  $\tilde{\pi}$ of the rotation $R_k$ , using
 (\ref{tensor prop lin}) and (\ref{tensor prop prod}), is
\[\tilde{\pi}(R_k) =\exp(\eta \varphi)=\exp ((I_2\otimes i\sum k_j\frac{1}{2}\sigma_j) \varphi)= I_2\otimes \exp( i\sum k_j\sigma_j\frac{\varphi}{2})\]
 for some constant $ \varphi$. Since $i(\sum k_j\sigma_j)^2=-I_2$, we get
 \begin{equation}\label{rep spin rotaion}
   \tilde{\pi}(R_k)= I_2\otimes\left(
  \begin{array}{cc}
      \cos \frac{\varphi}{2} + k_3 i\sin \frac{\varphi}{2} & (ik_1+k_2)\sin \frac{\varphi}{2} \\
        (ik_1- k_2)\sin \frac{\varphi}{2} & \cos \frac{\varphi}{2} -k_3 i\sin \frac{\varphi}{2} \\
              \end{array}
                   \right)
\end{equation}
for some constant $\varphi .$ Thus, we have a full description of the representation $\tilde{\pi}$ of the Lorentz group
on $M$.

To obtain the representation $\pi$ based on the usual electromagnetic tensor $F^\beta_\alpha$, we use
(\ref{Faraday real and compl}) expressing the connection of this tensor to the complex Faraday tensor. For boost
$B_k$ in direction $\mathbf{k}$, using (\ref{Faraday real and compl}) and (\ref{commuting f fbar}) we get
\begin{equation}\label{pi repr}
 \pi(B_k)=\exp\left((\xi+\xi^*)\psi \right )=
 \exp\left(\xi \psi\right )\exp\left(\xi^* \psi\right )=
 \tilde{\pi}(B_k)\tilde{\pi}^*(B_k).\end{equation}
 This defines the usual representation of a boost on the Minkowski space. Similarly, one can check that $\pi(R_k)$ defines the usual representation of a rotation on Minkowski space.

Note that all our representations are reducible. For representation $\pi$, the real and imaginary parts of $M$ are invariant. The operation of complex conjugation preserving $ReM$ and reversing $ImM$, So that the complex conjugate $y^*$ of a vector of $M$ is $y^*=Re\,y-iIm\,y$. This conjugation commutes with, and is invariant under, the representation $\pi$. Similarly, for representation $\tilde{\pi}$, from (\ref{rep spin boost}) and
(\ref{rep spin rotaion}) there are two invariant subspaces $M_1=span\{\mathbf{n}_0,\mathbf{n}_1\}$ and
 $M_2=span\{\mathbf{n}_2,\mathbf{n}_3\}$.  As a result, we introduce a \textit{conjugation}, denoted by $y^{\#}$, defined by the action of an operator $J$
       \begin{equation}\label{Jdef}
        J=\left(
         \begin{array}{cc}
           1 & 0 \\
           0 & -1 \\
         \end{array}
       \right) \otimes I_2,\;\Rightarrow\; (y^0,y^1,y^2,y^3)^{\#}=(y^0,y^1,-y^2,-y^3)
       \end{equation}
 which preserves $M_1$ and reverses $M_2$.  This conjugation replaces the complex conjugation for the representation $\pi$. It commutes with, and is invariant under, the representation  $\tilde{\pi}$. A similar conjugation exists also for the representation  $\tilde{\pi}^*$.

 \section{Lorentz invariant scale-free scalar associated with a null-vector}

\textit{\textbf{Claim }}Let $a$ be a null-vector with coordinates $a^\mu$ in the NP-basis. Then,
 the \textit{dimension-less constant} $\zeta=\frac{a^2}{a^0}$ is an invariant scalar with respect to representation
  $\tilde{\pi}$ on $M$.

 This constant coincides with the ``single complex parameter" occurring during the stereographic projection of the celestial sphere to the Agrand plane (see \cite{PenroseRindler} v.1 p.15).

 Let us first check that
$\frac{a^2}{a^0}$ is invariant under a boost $B_k$. By  (\ref{rep spin boost}), replacing $\frac{\psi}{2}$ with $\psi$, we have
\[\left(
    \begin{array}{c}
      a^{0'} \\
      a^{1'} \\
      a^{2'} \\
      a^{3'} \\
    \end{array}
  \right)=\tilde{\pi}(B_k)\left(
                     \begin{array}{c}
                       a^0 \\
                       a^1 \\
                       a^2 \\
                       a^3 \\
                     \end{array}
                   \right)=\left(
                             \begin{array}{c}
                                a^0(\cosh \psi + k_3\sinh \psi)+  a^1(k_1-ik_2)\sinh \psi\\
                                a^0(k_1+i k_2)\sinh \psi+  a^1(\cosh \psi -k_3\sinh \psi) \\
                                a^2(\cosh \psi + k_3\sinh \psi)+  a^3(k_1-ik_2)\sinh \psi\\
                                a^2(k_1+i k_2)\sinh \psi+  a^3(\cosh \psi -k_3\sinh \psi)  \\
                             \end{array}
                           \right),
                           \]
 implying, by the last equation of (\ref{Zero Int NP}), that
 \[\frac{a^{2'}}{a^{0'}}=\frac{a^2(\cosh \psi + k_3\sinh \psi)+  a^3(k_1-ik_2)\sinh \psi}
 {a^0(\cosh \psi + k_3\sinh \psi)+  a^1(k_1-ik_2)\sinh \psi}\]\[=
 \frac{a^2}{a^0}\,\frac{\cosh \psi + k_3\sinh \psi+ \frac{a^3}{a^2} (k_1-ik_2)\sinh \psi}
 {\cosh \psi + k_3\sinh \psi+ \frac{a^1}{a^0} (k_1-ik_2)\sinh \psi}
 = \frac{a^2}{a^0}\,.\]

 Let us now check that $\frac{a^0}{a^2}$ is invariant under a rotation $R_k$. By use of (\ref{rep spin rotaion}), replacing $\frac{\varphi}{2}$ with $\varphi$, we have
\[\left(
    \begin{array}{c}
      a^{0'} \\
      a^{1'} \\
      a^{2'} \\
      a^{3'} \\
    \end{array}
  \right)=\tilde{\pi}(R_k)\left(
                     \begin{array}{c}
                       a^0 \\
                       a^1 \\
                       a^2 \\
                       a^3 \\
                     \end{array}
                   \right)=\left(
                             \begin{array}{c}
                                        a^0(\cos \varphi + k_3 i\sin \varphi)+  a^1(ik_1+k_2)\sin \varphi\\
                                a^0(ik_1- k_2)\sin \varphi+  a^1(\cos \varphi -k_3 i\sin \varphi) \\
                                a^2(\cos \varphi + k_3 i\sin \varphi)+  a^3(ik_1+k_2)\sin \varphi\\
                                a^2(ik_1- k_2)\sin \varphi+  a^3(\cos \varphi -k_3 i\sin \varphi) \\
                             \end{array}
                           \right),
                           \]
 implying, by the last equation of (\ref{Zero Int NP}), that
 \[\frac{a^{2'}}{a^{0'}}=\frac{ a^2(\cos \varphi + k_3 i\sin \varphi)+  a^3(ik_1+k_2)\sin \varphi }
 { a^0(\cos \varphi + k_3 i\sin \varphi)+  a^1(ik_1+k_2)\sin \varphi}\]\[=
 \frac{a^2}{a^0}\frac{\cos \varphi + k_3 i\sin \varphi+  \frac{a^3}{a^2}(ik_1+k_2)\sin \varphi }
 { \cos \varphi + k_3 i\sin \varphi+ \frac{a^1}{a^0}(ik_1+k_2)\sin \varphi }
 = \frac{a^2}{a^0}\,.\]

 Thus, we have shown that the dimension-less constant $\zeta=\frac{a^2}{a^0}$ is invariant under the representation
  $\tilde{\pi}$ of the Lorentz transformations and thus is a scalar.

  Note that repeating a similar argument for representation $\tilde{\pi}^*$  based on the complex adjoint Faraday tensor, satisfying (\ref{Fraday adj in NP}), we get that $\zeta=\frac{a^2}{a^0}$ is not invariant under this representation. But the other pair of dimension-less constants from (\ref{Zero Int NP}), mainly $\frac{a^1}{a^0}=\frac{a^3}{a^2}$ is invariant under this representation. For the representation $\pi$ which by  (\ref{pi repr}) is the product of the
  representations $\tilde{\pi}$ and $\tilde{\pi}^*$  both constants will not be invariant.

  Let us return back to the usual basis $\{\mathbf{e}_\mu\}$ in space-time. Let $a$ be a null-vector on the positive light-cone. We will chose spherical coordinates in space. The in the usual basis $a$ is of the form
\begin{equation}\label{backward cone spherical}
    a=x^\mu=(r,r\cos\varphi\sin\theta,r\sin\varphi\sin\theta,r\cos\theta)=
    r(1,\cos\varphi\sin\theta,\sin\varphi\sin\theta,\cos\theta):=r\,n(\varphi,\theta).
\end{equation}
From (\ref{q in term r}) we get that our invariant dimension-less constant $\zeta=\frac{a^2}{a^0}$  in $\{\mathbf{e}_\mu\}$ basis is
\begin{equation}\label{z in spherical}
    \zeta=\frac{a^2}{a^0}=\frac{x^1-ix^2}{x^0+x^3}=
    \frac{\sin\theta}{1+\cos\theta}e^{-i\varphi}=\tan\frac{\theta}{2}e^{-i\varphi}.
\end{equation}

 \section{Scalar complex potential of the electromagnetic field generated by a moving charge}

In this section we introduce a complex scalar potential of the
electromagnetic field generated by a moving point charge $q$.

Denote by $P$ a point in space-time at which we want to calculate
the four potential, which we will call the observer. Denote by $f(\tau): \mathbb{R}\rightarrow \mathbb{R}^4$ the
world-line of the charge $q$ generating our electromagnetic field. Let the point
$Q=f(\tilde{\tau})$ be the unique point of intersection of the past light cone at $P$ with the
world-line $f(\tau)$ of the charge. We call $\tilde{\tau}$ the  the \textit{retarded time} of the potential.
 Note that radiation emitted at $Q$ at the retarded time will reach $P$ at time
$t$ corresponding to this point, and the potential at $P$ will depend on the position and the
velocity of the charge only at proper time $\tilde{\tau}$, see FIG.1.
\begin{figure}[h!]
  \centering
\scalebox{0.4}{\includegraphics{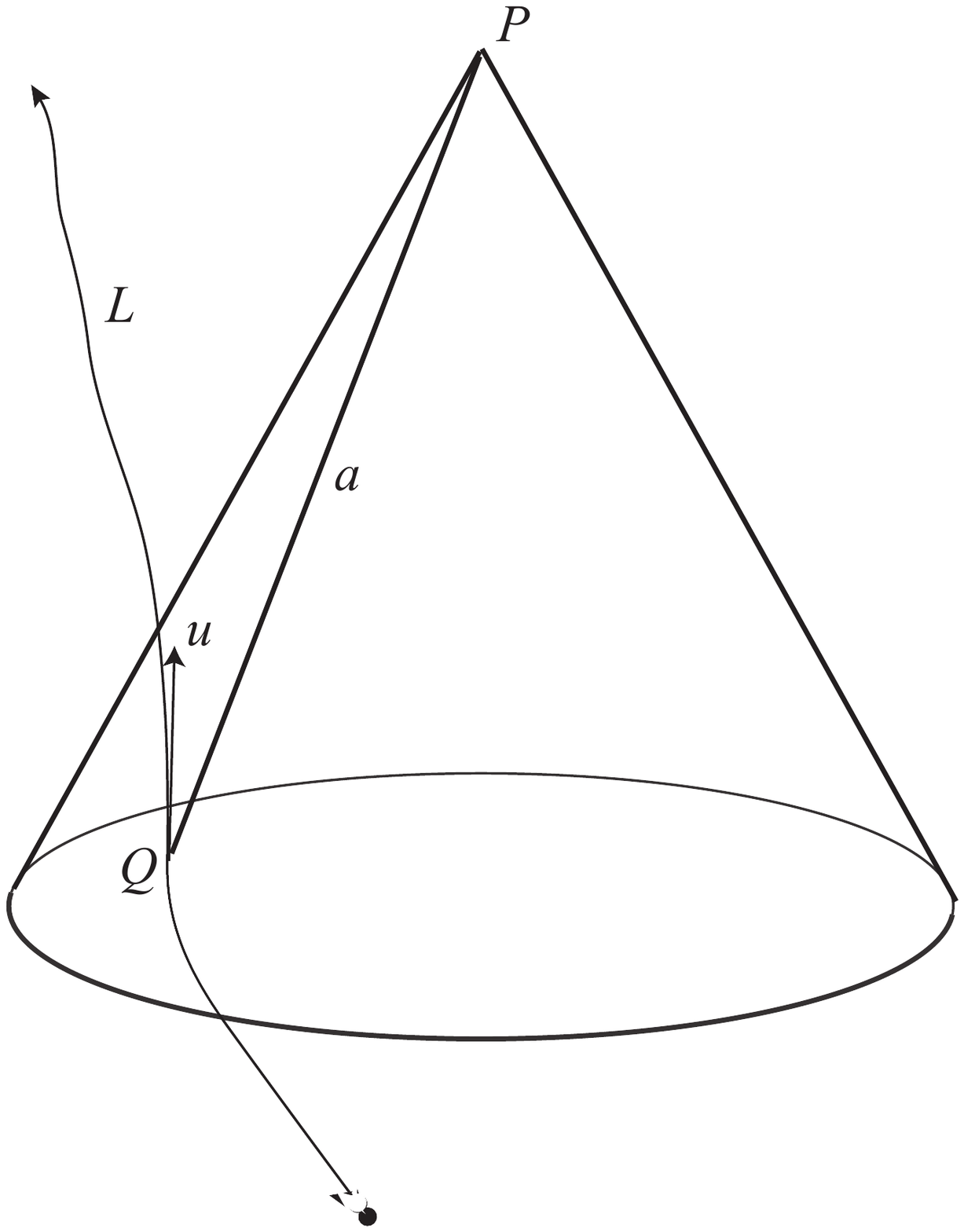}}
  \caption{The four-vectors associated with an observer and a moving charge.}\label{chargePotent}
\end{figure}

Let $K$ be an inertial reference frame in space-time with BT $\{\mathbf{n}_\mu\}.$
  and coordinates $(y^0,y^1,y^2,y^3).$ Denote by $y^\mu$ the  coordinates of $P$ in this basis and denote by $\tilde{y}^\mu$ the the  coordinates$Q$ of of the charge at the retarded time.
Introduce a 4-vector $a=\overrightarrow{QP}$. Its coordinates in the BT are
\begin{equation}\label{a def}
 a^\mu=y^\mu-\tilde{y}^\mu\,.
\end{equation}
This vector is a null (light-like) vector in space-time and thus satisfies (\ref{Zero Int NP}).
Denote by $w=w(P)=w^\nu$ the 4-velocity of the charge at the point $P$ with coordinates
expressed in BT, corresponding to time $\tilde{\tau}$. Note that only two 4-vectors:
the null-vector $a$ defining the relative position of the charge at retarded time with respect
to the point of observation of the potential, and the  time-like vector $w$ of
length 1 describing the 4-velocity of the charge at the retarded time, are independent of the choice
of the reference frame.

 From these two vectors we can generate a co-variant scalar $a\cdot w$. From the previous section, we know
that there is also a dimension-less scalar $\zeta=\frac{a^2}{a^0}=\frac{a^3}{a^1}$ invariant under the
representation $\tilde{\pi}$. We want the scalar
potential to be a function of this dimension-less scalar.

To identify the function of the scalar potential, note that the electric force depends on
the distance from the charge as $\frac{1}{r^2}$ and, as it was explained in the Introduction,
the force has to be a second derivative of the potential. So, the natural candidate for the
function of the scalar potential is a function proportional to $\ln $.

\noindent\textbf{Definition} We define a\textit{ scalar complex  potential} $S (P)$ at the observer point
$P$ of a moving charge $q$ by
\begin{equation}\label{WitPotin_gen}
    S (P)=q\ln\zeta,
\end{equation}
where $\zeta$ is the dimension-less scalar associated with the null vector $a=\overrightarrow{QP}$ connecting
the position of the charge $Q$ at retarded time with the observer point. In BT this potential is equal
\begin{equation}\label{WitPotinNP}
    S (P)=q\ln\left(\frac{a^2}{a^0}\right)=q(\ln a^2-\ln a^0)=q\ln\left(\frac{a^3}{a^1}\right)=
    q(\ln a^3-\ln a^1),
\end{equation}
with $a^\mu$ defined by (\ref{a def}). For the usual basis, from (\ref{z in spherical}) this potential
is
\begin{equation}\label{WitPotinRB}
    S (P)=q\ln\zeta=q\ln\frac{x^1-ix^2}{x^0+x^3}=q\ln (\tan\frac{\theta}{2})-iq\varphi,
\end{equation}
where $\varphi,\theta$, as in (\ref{backward cone spherical}) are the angles in spherical coordinates of the space component of the vector
$a=\overrightarrow{QP}$.

Our complex scalar potential $S $ is a combination $F-iG$ of the two scalar potentials
introduced by Whittaker, see  \cite{Whittaker} and \cite{Ruse}.

\section{The 4-potential and the electromagnetic field of a moving charge}

 For future calculations, we will need a formula for partial
derivatives of $a$ and the retarded time $\tilde{\tau}$ with respect to
$y^\mu$, for any given $\mu$. Denote by $\check{P}=P+\Delta ^\mu P$ the point with
coordinates $y+dy^\mu$ and by $\check{Q}=Q+\Delta ^\mu Q$ the position
of the charge at the corresponding retarded time. To first
order in $dy^\mu$, we can write $\Delta ^\mu Q= d\tilde{\tau} w^\mu$ for some
 $d\tilde{\tau}.$ Then the 4-vector $\check{a}=\overrightarrow{\check{Q}\check{P}}$
 is approximately $\check{a}=a+dy^\mu-d\tilde{\tau} w^\mu$. Since
$\check{a}$ and $a$ are null, we get, to first order,
\[ (\check{a})^2=0 \;\Rightarrow (a)^2 -2\tilde{\eta}_{\lambda\nu}a^\lambda w^\nu d\tilde{\tau}+2\tilde{\eta}_{\mu\nu}a^\nu dy^\mu =0
\;\Rightarrow \tilde{\eta}_{\mu\nu}a^\nu dy^\mu =(a\cdot w )d\tilde{\tau}.\] Thus,
\begin{equation}\label{DerRetadedTime}
\frac{\partial\tilde{\tau}}{\partial y^\mu}=\frac{\tilde{\eta}_{\mu\nu}a^\nu}{a\cdot w}=\frac{a_\mu}{a\cdot w}.\end{equation}

From (\ref{a def}), the component $\nu$ of $a$ is
$a^\nu=y^\nu-\tilde{y}^\nu$. Thus,
\begin{equation}\label{derivNPbasis}
 \frac{\partial a^\nu}{\partial y^\mu}=\delta_\mu^\nu-\frac{\partial \tilde{y}^\nu}{\partial y^\mu}
  =\delta_\mu^\nu-\frac{\partial \tilde{y}^\nu}{\partial \tilde{\tau}}
   \frac{\partial \tilde{\tau}}{\partial y^\mu}
  =\delta_\mu^\nu- \frac{w^\nu a_\mu}
{a\cdot w }\,.\end{equation}
This gives
\[\frac{\partial (a \cdot w)^{-1}}{\partial y^\mu}=-(a \cdot w)^{-2}\frac{\partial (a \cdot w)}{\partial y^\mu}
=-(a \cdot w)^{-2}\frac{\partial (\tilde{\eta}_{\lambda\nu}w^\lambda a^\nu)}{\partial y^\mu}
=-(a \cdot w)^{-2}\tilde{\eta}_{\lambda\nu}\left(a^\nu\frac{\partial w^\lambda }{\partial y^\mu}+
w^\lambda\frac{\partial  a^\nu}{\partial y^\mu}\right).\]
We assume now that $\frac{\partial w^\lambda }{\partial y^\mu}=0$, excluding accelerating charges.
Then, by use of (\ref{derivNPbasis}) and  the fact that $(w)^2=\tilde{\eta}_{ lk}w^lw^k=1$, we get
\begin{equation}\label{derivScalProdNPbasis}
 \frac{\partial (a \cdot w)^{-1}}{\partial y^\mu}=-(a \cdot w)^{-2}\tilde{\eta}_{\lambda\nu}w^\lambda
( \delta_\mu^\nu-\frac{w^\nu a_\mu}{a \cdot w }) =(a \cdot w)^{-2}(
\frac{ a_\mu}{a \cdot w }-w_\mu )= \frac{ a_\mu}{(a \cdot w)^3}
 -\frac{ w_\mu}{(a \cdot w)^2}.
\end{equation}

We define a \textit{complex 4-potential} $\tilde{A}_\mu$  to be the conjugate to the gradient of the scalar complex potential vector. By use of (\ref{Jdef}) we have
\begin{equation}\label{4potential}
   \tilde{A}_\mu=-J_\mu^\nu\frac{\partial S }{\partial y^\nu}=-J_\mu^\nu\tilde{\partial}_\nu S =
   (-\frac{\partial S }{\partial y^0},
-\frac{\partial S }{\partial y^1},\frac{\partial S }{\partial y^2},
\frac{\partial S }{\partial y^3})\,.
\end{equation}

Using (\ref{derivNPbasis}), we get:
\[\tilde{A}_0=\frac{\partial S }{\partial y^0}=\frac{\partial q( \ln a^3-\ln a^1)}{\partial y^0}
=\frac{q}{a^3}\,\frac{w^3a_0}{a \cdot w}-\frac{q}{a^1}\,\frac{w^1a_0}{a \cdot w}=
\frac{q}{a \cdot w}\left(w_0-a_0\frac{w^1}{a^1}\right),\]
\[\tilde{A}_1=\frac{\partial S }{\partial y^1}=\frac{\partial q(\ln a^2-\ln a^0)}{\partial y^1}
=\frac{q}{a^2}\,\frac{w^2a_1}{a \cdot w}-\frac{q}{a^0}\,\frac{w^0a_1}{a \cdot w}=
\frac{q}{a \cdot w}\left(w_1-a_1\frac{w^0}{a^0}\right),\]
\[\tilde{A}_2=-\frac{\partial S }{\partial y^2}=\frac{\partial q(\ln a^1-\ln a^3)}{\partial y^2}
=\frac{q}{a^1}\,\frac{w^1a_2}{a \cdot w}-\frac{q}{a^3}\,\frac{w^3a_2}{a \cdot w}=
\frac{q}{a \cdot w}\left(w_2-a_2\frac{w^3}{a^3}\right),\]
\[\tilde{A}_3=-\frac{\partial S }{\partial y^3}=\frac{\partial q(\ln a^0-\ln a^2)}{\partial y^3}
=\frac{q}{a^0}\,\frac{w^0a_3}{a \cdot w}-\frac{q}{a^2}\,\frac{w^2a_3}{a \cdot w}=
\frac{q}{a \cdot w}\left(w_3-a_3\frac{w^2}{a^2}\right).\]

\section{The electromagnetic field of a moving charge}
By substituting (\ref{4potential}) into (\ref{Fj fromA NP}) we get
\[F_j=-\tilde{\partial}^\nu (I_2\otimes \sigma_j)^{\mu}_\nu \tilde{A}_\mu=
\tilde{\partial}^\nu (I_2\otimes \sigma_j)^{\mu}_\nu J_\mu^\lambda\tilde{\partial}_\lambda S .\]
Since
\[(I_2\otimes \sigma_j)^{\mu}_\nu J_\mu^\lambda=\left(
                                                  \begin{array}{cc}
                                                    \sigma_j & 0 \\
                                                    0 & -\sigma_j \\
                                                  \end{array}
                                                \right)=(\alpha_j)_\nu^\lambda \,,\]
 where $(\alpha_j)$ are the known $\alpha$-matrices of Dirac, (see \cite{LL4}) we have:
 \begin{equation}\label{F as from Fhi}
   F_j=\tilde{\partial}^\nu (\alpha_j)_\nu^\lambda\tilde{\partial}_\lambda S .
 \end{equation}
This give explicit formulas for each component of the $F$:
\[F_1=\frac{\partial ^2S }{\partial y^1\partial y^3}-
\frac{\partial ^2S }{\partial y^0\partial y^2},\;\;
F_2=i\left(\frac{\partial ^2S }{\partial y^1\partial y^3}+
\frac{\partial ^2S }{\partial y^0\partial y^2}\right)\,,\]
\[F_3=\frac{\partial ^2S }{\partial y^1\partial y^2}+
\frac{\partial ^2S }{\partial y^0\partial y^3}\,.\]

By differentiating the first derivatives of $S$ and applying (\ref{derivNPbasis}) and (\ref{derivScalProdNPbasis}), for the potential of a moving charge we get
\[ \frac{\partial ^2S }{\partial y^1\partial y^3}=q\frac{w^0a_1+w^2a_3}{(a \cdot w)^3},\;\;\;\;
\frac{\partial ^2S }{\partial y^0\partial y^2}=-q\frac{w^1a_0+w^3a_2}{(a \cdot w)^3},\]
and
\[\frac{\partial ^2S }{\partial y^1\partial y^2}=\frac{q}{2(a \cdot w)^2}-
q\frac{w^3a_3+w^1a_1}{(a \cdot w)^3}\]
\[
\frac{\partial ^2S }{\partial y^0\partial y^3}=\frac{-q}{2(a \cdot w)^2}
+q\frac{w^0a_0+w^2a_2}{(a \cdot w)^3}\,.\]
This defines a formula for the Faraday vector of the electromagnetic field of a moving charge
\begin{equation}\label{Fjfinal}
    F_j=q\frac{a_\mu (\tilde{K}_j)_\nu^\mu w^\nu}{(a \cdot w)^3}\,,
\end{equation}
where $(\tilde{K}_j)_\nu^\mu$ is defined by (\ref{KjinNP}).
 
In the  basis $\mathbf{e}_\mu$ this formula becomes
\begin{equation}\label{FjfinalRgular}
    F_j=q\frac{x_0u_j-x_ju_0+i\varepsilon _j^{kl}( x_k u_l)}{(x \cdot u)^3}\,,
\end{equation}
 This formula coincides  with the usual formula for the field of a moving charge (see, for example, \cite{Jackson} p. 573 ).

Direct calculation shows that $S $ satisfies the wave equation
\begin{equation}\label{wave}
    \square ^2 S  =\tilde{\partial}^\mu \tilde{\partial}_\mu=
    2\left(\frac{\partial ^2S }{\partial y^0\partial y^3} -
\frac{\partial ^2S }{\partial y^1\partial y^2}\right)=0\,.
\end{equation}
For the 4-potential (\ref{4potential}) this implies that
\begin{equation}\label{waveAmu}
    \frac{\partial \tilde{A}_3}{\partial y^0}-\frac{\partial \tilde{A}_1}{\partial y^2}=0,\;\;
     \frac{\partial \tilde{A}_0}{\partial y^3}-\frac{\partial \tilde{A}_2}{\partial y^1}=0\,.
\end{equation}

\section{Scalar complex potential of an electromagnetic field}

Any electromagnetic field is generated by a collection of moving charges. We may
assume that charges close to each other move with velocities that do not vary significantly.
The sources of the electromagnetic field may be represented by the charge densities $\rho (x)$
on space-time 4-vector $x$. We assume that the potential depends additively on the charges generating the field.
By choosing a BT in space-time, the scalar complex potential of the electromagnetic field is given by
\begin{equation}\label{scalPotGenera}
    S(y) =\int_{S^-(y)}\ln(\frac{a^2}{a^0})\rho(y+a)da,
\end{equation}
where $S^-(y)$ denotes the backward light-cone at $y$.

Let us return back to the usual basis
$\{\mathbf{e}_\mu\}$ in space-time. If we use spherical coordinates in space as in (\ref{backward cone spherical})
and the formula (\ref{WitPotinRB}) for the scalar potential of a charge, then we can rewrite
(\ref{scalPotGenera}) as
\begin{equation}\label{wave of field}
     S(x) =\int_0^{2\pi} d\varphi \int_0^\pi \ln (\tan\frac{\theta}{2})\sin\theta d\theta
     \int_0^\infty \rho(x+r\,n(\varphi,\theta))r^2dr-i\int_0^{2\pi} \varphi d\varphi \int_0^\pi \sin\theta
     d\theta\int_0^\infty \rho(x+r\,n(\varphi,\theta))r^2dr\;.
\end{equation}

Since equation (\ref{wave}) is a linear homogeneous equation it will hold for for a scalar potential of arbitrary
electromagnetic field. Also equations (\ref{waveAmu}) will hold for such a field.
Using these equations and (\ref{tensorfroAmuNP}) we get obtain the following formula for the connection
of the 4-potential and Faraday tensor
\begin{equation}\label{F fromA in NP}
  \widetilde{\mathcal{F}}_{\alpha\beta}=\left(
                                          \begin{array}{cc}
                                            0 & 1 \\
                                            -1 & 0 \\
                                          \end{array}
                                        \right)
  \otimes \left(
   \begin{array}{cc}
   \left( \frac{\partial \tilde{A}_0}{\partial y^2}-\frac{\partial \tilde{A}_2}{\partial y^0}\right) &
    \left(\frac{\partial \tilde{A}_3}{\partial y^0}-\frac{\partial \tilde{A}_0}{\partial y^3}\right) \\
    \left( \frac{\partial \tilde{A}_2}{\partial y^1}-\frac{\partial \tilde{A}_1}{\partial y^2}\right)  &
    \left(\frac{\partial \tilde{A}_1}{\partial y^3}-\frac{\partial \tilde{A}_3}{\partial y^1}\right) \\
                                                    \end{array}
                                                  \right)=(-1)^{(\alpha +\beta)} \left( \frac{\partial \tilde{A}_\alpha}{\partial y^\beta}-\frac{\partial \tilde{A}_\beta}{\partial y^\alpha}\right)\,.
\end{equation}
This show that our differential operator $\widetilde{\nabla}_c\times $ in (\ref{tensorfroAmuNP}) is
\begin{equation}\label{curl complex}
   (\widetilde{\nabla}_c\times)^\mu_{\alpha\beta} =(-1)^{(\alpha +\beta)}\left( \delta ^\mu _\alpha \partial _\beta- \delta^\mu _\beta \partial _\alpha\right),
\end{equation}
which is close to the  $curl=({\nabla}\times)^\mu_{\alpha\beta} = \delta^\mu _\alpha\partial _\beta- \delta^\mu _\beta \partial _\alpha$.
Note that without conjugation in definition of  $\tilde{A}^\mu$, we would get a zero answer for $\widetilde{\mathcal{F}}_{\alpha\beta}$. This corresponds to the known identity $\nabla\times\nabla S=0$ for any scalar function $S$.

\section{Discussion and conclusion}

We have shown that the E. T. Whittaker scalar potentials of a moving charge became one complex potential
(\ref{WitPotinRB})on space-time. Thus, any electromagnetic field can be described by use of the complex scalar
potential which is as the wave-function of Quantum Mechanics, a complex-valued function on space-time. 
We defined a representation $\tilde{\pi}$ of the Lorentz group on the complexified Minkowski space $M$ and its adjoint one $\tilde{\pi}^*$, which commutes with it. The usual representation $\pi$ of the Lorentz group on $M$ is the product of representations $\tilde{\pi}$ and $\tilde{\pi}^*$. We defined a Lorentz Invariant
conjugation (\ref{Jdef}) on $M$ under the representation $\tilde{\pi}$.

For any electromagnetic field  by use of (\ref{wave of field}) we obtain a complex-valued function,
that we called the complex scalar potential,
which is defined by the distribution of the charges generating the field. The components of
the Faraday vector, describing the electro-magnetic force of the field, may be obtained from the potential
by (\ref{F as from Fhi}) as
\[   F_j=\tilde{\partial}^\nu (\alpha_j)_\nu^\lambda\tilde{\partial}_\lambda S .\]
where $(\alpha_j)$ are the known $\alpha$-matrices of Dirac. This ones more reveals the connection between our 
description of the classical electromagnetic field and Quantum Mechanics.
 From the Claim in Section 6, it follows that the complex scalar potential is an invariant scalar under
the representation $\tilde{\pi}$.

 We show that the Faraday tensor $\widetilde{\mathcal{F}}$ is the complexified $curl$ defined by (\ref{curl complex}) of the conjugate of the gradient of the complex potential, which can be written as.
  \begin{equation}\label{Faraday tensor from scalar potential}
    \widetilde{\mathcal{F}}=\widetilde{\nabla}_c\times (\widetilde{\nabla}S)^\#\,.
 \end{equation}
 The real part of
the Faraday tensor $\mathcal{F}$ coincides with the usual
electromagnetic tensor of the field. We obtain an explicit formula for
the Faraday vector (\ref{FjfinalRgular}) of a moving charge.
We have shown that the complex scalar potential satisfies the wave equation (\ref{wave}).

Note that the complex logarithm is not defined uniquely. So,the scalar complex potential is a 
multi-valued function. The Aharonov-Bohm effect \cite{AharonovBohm59} revealed that in presence of an electromagnetic field the wave function of a particle is multiplied by a multi-valued function defined
by the field. This is similar to our observation.

\medskip

The first author would like to thank Prof. Donald  Lynden-Bell for  bringing to his attention the result of
E. T. Whittaker and providing him with a copy of  Whittaker's paper. Both would like to thank Prof. U. Sandler,
Dr. Y. Itin, Dr. V. Ostapenko and Dr. Z. Scaar for helpful discussions and remarks.

\end{document}